\documentclass[review,12pt]{elsarticle}
\usepackage{lineno,hyperref}
\modulolinenumbers[5]

\journal{Journal of \LaTeX\ Templates}

\usepackage{amsmath}
\usepackage{graphicx}
\title{Acoustoelectric Effect in  degenerate Carbon Nanotube}

\author[rvt]{K. A. Dompreh\corref{cor1}\fnref{fn1}}
\author[focal]{N. G. Mensah}
\author[rvt]{S. Y. Mensah}
\author[rvt]{F. Sam}
\author[rvt]{ A.K. Twum}
\address[rvt]{Department of Physics, College of Agriculture and Natural Sciences, U.C.C, Ghana.}
\address[focal]{Department of Mathematics, College of Agriculture and Natural Sciences, U.C.C, Ghana}
\cortext[cor1]{Corresponding author\\ Email: kwadwo.dompreh@ucc.edu.gh}
\ead[url]{kwadwo.dompreh@ucc.edu.gh}

\date{}

\begin{document}
\begin{abstract}
\noindent 
Acoustoelectric Effect $AE$ in degenerate Carbon Nanotube ($CNT$) was theoretically studied for hypersound in the 
regime $ql >> 1$. The dependence of acoustoelectric current $j^{ac}$ on the acoustic wave 
number $\vec{q}$ and frequency $\omega_q$ at $T = 10K$ and scattering angle ($\theta > 0$) 
was evaluated at various harmonics $n =\pm 1, 2, ...$ (where $n$ is an integer). In the first 
harmonics ($n = \pm 1$), the non-linear dependence of $j^{ac}$ on $\omega_q$ and $\vec{q}$ 
were obtained. For $n = \pm 2$, the numerically evaluated $j^{ac}$  qualitatively agreed with an experimentally 
obtained result. 
\end{abstract}

\maketitle
\section*{Introduction}
Carbon Nanotubes  (CNT) discovered by Iijima ~\cite{1} are emerging as important materials for electronic applications 
due to their remarkable electrical and mechanical properties~\cite{2,3,4}. The metallic and semiconducting 
Single-Walled Carbon Nanotube (SWCNT)  have been proposed as the most viable materials to  
develop high performance thin films to completely eliminate the use of critical metals in electronic devices such as: 
i) Indium in transparent conducting films ($TCF$, indium oxide doped by tin, $ITO$ ) and ii) Indium and Gallium as semiconductor
${In–Ga–Zn–O}$ ($a-IGZO$) in thin film field effect transistors ($TFTs$) for applications in optoelectronics~\cite{5,6,7,8, 9}. 
The unusual band structure~\cite{10,11,12} of $CNT$, coupled with large electron densities and high drift velocities 
(with electron mobility of $\mu\approx 10^5 cm^2/Vs$) at room temperatures opens a way for employing carrier control 
processes rather than direct electrical control~\cite{13}. In the linear regime, electron-phonon interactions in $CNT$ 
at low temperatures leads to emission or absorption of large number of coherent acoustic phonons ~\cite{14,15,16}.
When the momentum of the acoustic phonons is absorbed by the conducting electrons, it leads to the 
appearance of $d.c$ electric field ~\cite{17,18,19,20,21}. This phenomena is known as Acoustoelectric effect $AE$~\cite{22,23,24,25,26}.  
Studies of Acoustoelectric effect in bulk materials such as Gallium Nitride ($GaN$)~\cite{27,28,29} with applications in  
$GaN$ film bulk acoustic resonators ($FBARs$), Indium Antimonide ~\cite{30} and $GaAs/LiNbO_3$~\cite{31} has being reported.

With the advent of low-dimensional materials, $AE$ has been intensively studied in quantum wells~\cite{32} to produce 
quantized current in $1-D$ channels~\cite{33} for light storage  and to induce charge pumping in nanotube quantum dots~\cite{34}.
$AE$ studies in superlattices~\cite{35,36}, quantum wires~\cite{37,38} and Zinc Oxide ($ZnO$) Nanowires ~\cite{39} has been reported. 
Recently, the need for acoustically driven current flow in semiconductor nano-
structures has received particular attention as a means of generating or controlling single electrons and 
photons for quantum information processing ~\cite{40,41,42}. 
From the hypersound absorption~\cite{43} studies conducted on $CNT$, it was found that $CNT$ exhibit good AE effect but till date, there is no 
general analytical theory of Acoustoelectric effect in $CNT$. However, a few  experimental work by 
Ebbecke et. al.~\cite{44} and   Reulet et. al~\cite{45}   in $CNT$ has been carried out.
In this paper the theoretical treatment of $AE$ in $CNT$ in the hypersound regime $ql >> 1$ (where $q$ is the acoustic wave number,
$l$ is the mean free path of an electron) is carried out. The general expression obtained is analysed numerically for harmonics
$n = \pm 1,\pm 2$ (where $n$ is an integer). The paper is organised as follows: In section $2$, the kinetic theory
based on the linear approximation for the phonon distribution function is setup, where, the rate of growth of the 
phonon distribution is deduced  and the acoustoelectric current ($j^{ac}$) is obtained. 
In section $3$,  the final equation is analysed numerically in a graphical form  at various 
harmonics. Lastly the  conclusion is presented in section $4$.  

\section{Theory} 
Proceeding from ~\cite{22,23}, the Acoustoelectric
current $j^{ac}$ in the hypersound regime $ql >> 1$ is given  as 
\begin{equation} 
j =-\frac{4\pi\tau e}{(2\pi)^3}{\vert C_q\vert} ^2 \int_0^{\infty}{v_i[f(p+q) - f(p)]\delta(\varepsilon(p+q)-\varepsilon(p)-\hbar\omega)d^3p}
\end{equation}
where the velocity $v_i = v(p+q) - v(p)$, $f(\varepsilon(p))$ is the distribution function, $p$ is the momentum of electrons
and $\tau$ is the relaxation constant. The linear energy dispersion $\varepsilon(\vec{p})$ relation 
for the $CNT$ is given as~\cite{46}
\begin{equation} 
\varepsilon(\vec{p}) = \varepsilon_0 \pm \frac{\sqrt{3}}{2\hbar}\gamma_0 b(\vec{p} - \vec{p}_0) \label{Eq_2}
\end{equation}
The $\varepsilon_0$ is the electron energy in the Brillouin zone at momentum $p_0$, $b$ is the 
lattice constant , $\gamma_0$ is the tight binding 
overlap integral ($\gamma_0 = 2.54$eV). The $\pm$ sign indicates that in the vicinity of the tangent point,
the bands exhibit mirror symmetry with respect to each point. After collision, $\vec{p^\prime}= (\vec{p}+\hbar \vec{q})cos(\theta)$
is the component directed along the CNT axis.  Where $\theta$ is the scattering angle. At low temperature 
($kT << 1$), the Fermi-Dirac equilibrium function is given as 
\begin{equation}
f_{\vec{p}} = exp(-\beta(\varepsilon_{\vec{p}} -\mu)) \label{Eq_3}
\end{equation}
with $\mu$ being the chemical potential, $\beta = 1/kT$, $k$ is the Boltzmann constant.
Inserting Eqn.($2$ and $3$) into  Eqn.($1$), and after some cumbersome calculations yield
\begin{equation}
j^{ac} = \frac{2e\pi\vert C_{\vec{q}}\vert^2} {\hbar} exp(-\beta(\varepsilon_0 -\chi \vec{p}_0))
 \{exp(-\beta\chi(\eta +\hbar \vec{q})cos(\theta))- exp(-\beta\chi\eta)\}\label{Eq_4}
\end{equation}
where $\chi = {\sqrt{3}\gamma_0 b}/{2\hbar}$, and 
$$\eta =\frac{-2{\hbar}^2\omega_{\vec{q}} + \gamma_0 b\sqrt{3}\hbar \vec{q} cos(\theta)}{\gamma_0 b\sqrt{3}(cos(\theta)-1)}$$ 
For acoustic phonons, $\vert{C_{\vec{q}}}\vert = \sqrt{{\Lambda^2 \hbar \vec{q}}/{2\rho V_s}}$.
Where $\Lambda $ is the deformation potential constant and $\rho$ is the density of the material.
Taking $\varepsilon_0 = \vec{p}_0 = 0$, the Eqn.($4$) finally reduces to  
\begin{equation}
j^{ac} =\frac{2e\vert {\Lambda}\vert^2\tau \hbar q^2 exp(-\beta\chi\eta)}{2\pi{\hbar\omega_q}} 
\{\sum_{n=-\infty}^{\infty}{\frac{exp(-n(\theta+\beta\chi\eta))}{I_n(\beta\chi(\eta +\hbar \vec{q}))} - 1}\} \label{Eq_5}
\end{equation}
where $I_n(x)$ is the modified Bessel function. 

\section*{Numerical analysis}
The analytical solution of Eqn($5$) is obtained numerically and the results  presented graphically. 
The parameters used in the numerical evaluation  are: $\vert {\Lambda}\vert = 9$eV, $b = 1.42$nm,
$q = 10^7$ cm$^{-1}$, $\omega_q = 10^{12}$s$^{-1}$,$V_s = 4.7\times10^5$ cm s$^{-1}$, $T = 10K$, and $\theta > 0$. 
The  dependence of $j^{ac}$ on the acoustic wave number ($\vec{q}$) and the frequency ($\omega_q$)
at various harmonics ($n = \pm 1,\pm 2$) are presented below.  For $n = \pm 1$,  the non-linear graph (with an initial curve)  
increases sharply  to a maximum  then  decreases to a constant minimum value (see Figure $1a$). 
By increasing the  values of $\vec{q}$, the graph shift to the right with decreasing amplitude.
In Figure $1b$, it was observed that at $\omega_q = 0.6\times10^{12}s$,
the $j^{ac}$ increases to a maximum point and then falls to a minimum value. Increasing the values of $\omega_q$, 
shift the graph to the right. More interesting is the nature of the acoustoelectric current $j^{ac}$. At $\omega_q = 0.6\times10^{12}s$,
the ratio of the  peaks balances on both side of the $j^{ac}$ axis. 
At $\omega_q = 0.65\times 10^{12}s$, the ratio of the $j^{ac}$ peaks is more towards the negative side but a reverse 
occurs when $\omega_q = 0.7\times 10^{12}s$. For $n = \pm 2$ (see Figure $2$ ($a$ and $b$)), 
the graph obtained  for $j^{ac}$ versus $\omega_q$ qualitatively agreed with an experimentally obtained results. 
In Figure $2b$, it is also observed that the dependence of $j^{ac}$ on $q$ is strongly non-linear. 
The ratio of $j^{ac}\over \Gamma$ (where $\Gamma$ is the hypersound absorption) in the absence of a drift velocity 
$V_D$ ~\cite{43} is given  as 
\begin{equation}
{j^{ac}\over\Gamma} = \frac{2e\tau\gamma_0 b \sqrt{3}}{\hbar}(cos(\theta) - 1)
\end{equation}
which is the Weinreich relation~\cite{21} and is dependent on the scattering angel $\theta$.  
For better understanding of the obtained graphs, a $3D$ graph of $j^{ac}$ versus $\omega_q$ and $\vec{q}$ 
are presented. For $n = \pm 1$, the ratio of the height of the peaks in the positive 
side of $j^{ac}$ far exceeded that in the negative side (see Figure $3a$). In Figure $3b$, for $n = \pm 2$, the graph showed
peaks at certain intervals.   
\begin{figure}[htp]
\includegraphics[width =7.0cm]{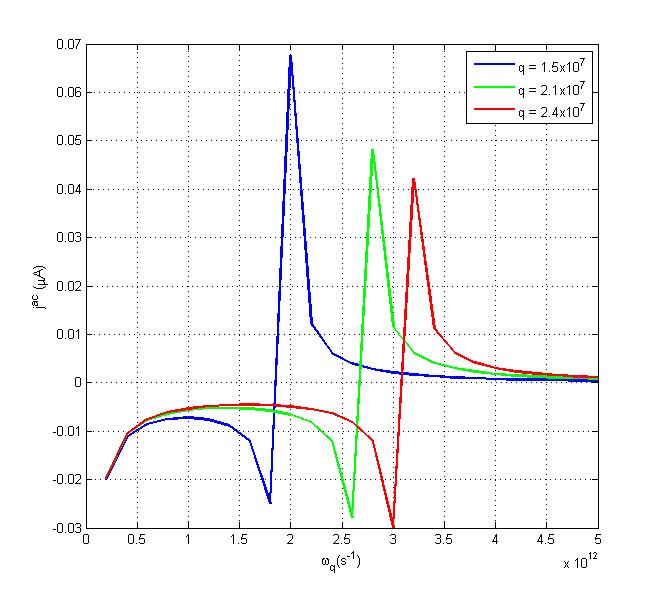}
\includegraphics[width =7.0cm]{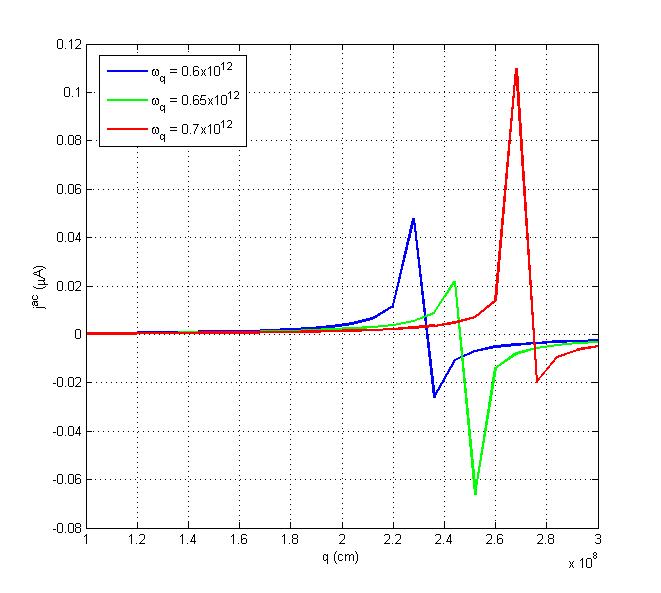}
\caption{  (a) Dependence of  $j^{ac}$ on $\omega_q$ at varying $\vec{q}$ (b) Dependence of $j^{ac}$ on $\vec{q}$
for varying $\omega_q$}
\end{figure}
\begin{figure}
\includegraphics[width =7.0cm]{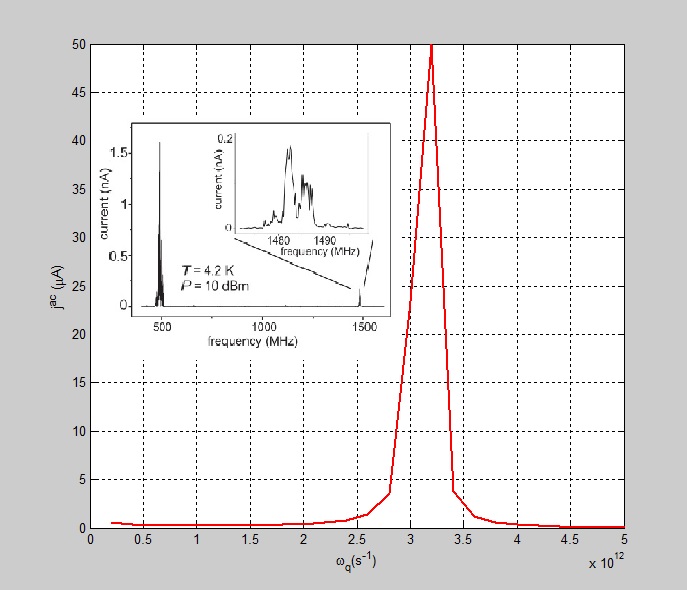}
\includegraphics[width =7.0cm]{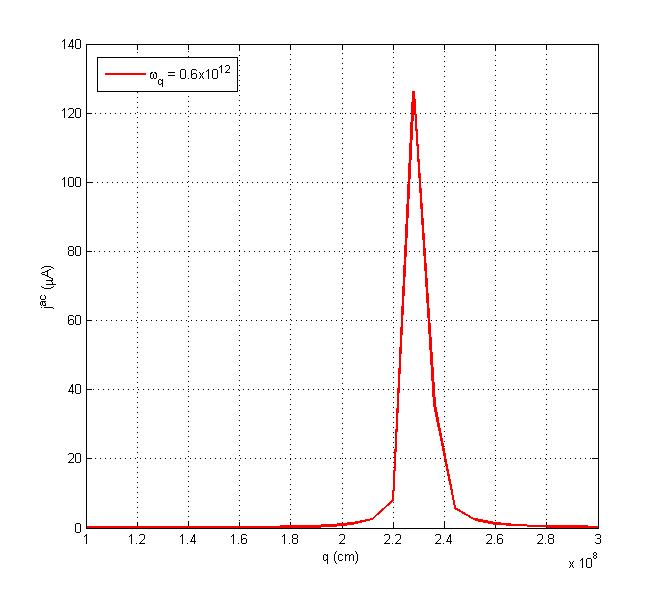}
\caption{  (a) Dependence of  $j^{ac}$ on $\omega_q$ at various $\vec{q}$ insert shows the 
experimentally obtained acoustoelectric current~\cite{44}
(b) Dependence of $j^{ac}$ on $\vec{q}$ for varying $\omega_q$. }
\end{figure}
\begin{figure}
\includegraphics[width = 7.0cm]{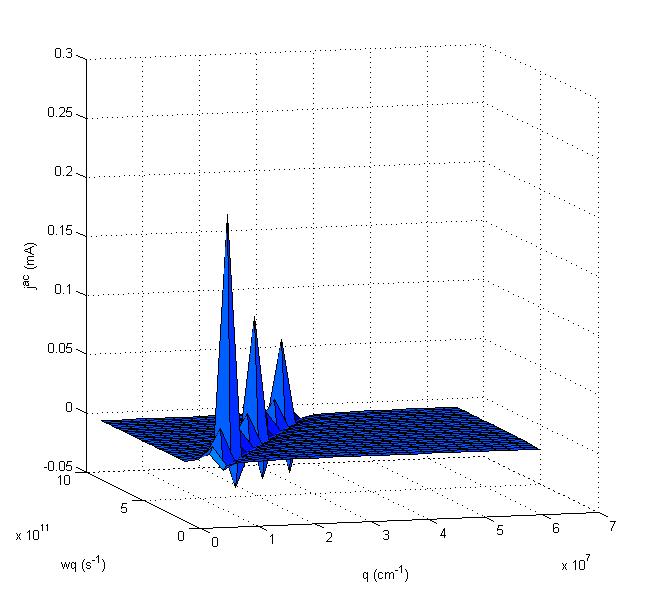}
\includegraphics[width = 7.0cm]{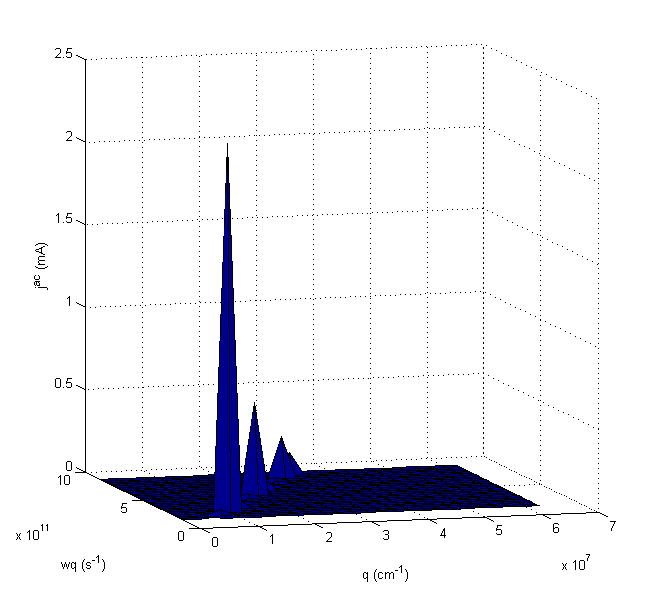}
\caption{   Dependence of  $j^{ac}$ on $\gamma$ and $\vec{q}$ at (a) the first harmonics $n = \pm 1$ 
(b) the second harmonics $n = \pm 2$} 
\end{figure}

\section{Conclusion}
The Acoustoelectric Effect $AE$ in a degenerate Carbon Nanotube $CNT$ is studied for hypersound in the 
regime $ql >> 1$.   A strong nonlinear dependence of $j^{ac}$ on the acoustic wavenumber $\vec{q}$ 
and the frequency $\omega_q$ are observed. The dominant mechanism for such non-linear behaviour is 
the Acoustoelectric Effect which give rise to the acoustoelectric current $j^{ac}$. The analytically 
obtained  acoustoelectric current $j^{ac}$ qualitatively agrees with an experimentally obtained result.

\renewcommand\refname{Bibliography}

\end{document}